\documentclass[conference]{IEEEtran}
\makeatletter
\def\ps@IEEEtitlepagestyle{%
  \def\@oddfoot{\mycopyrightnotice}%
  \def\@oddhead{\hbox{}\@IEEEheaderstyle\leftmark\hfil\thepage}\relax
  \def\@evenhead{\@IEEEheaderstyle\thepage\hfil\leftmark\hbox{}}\relax
  \def\@evenfoot{}%
}
\def\mycopyrightnotice{%
  \begin{minipage}{\textwidth}
  \scriptsize
  \copyright~2024 IEEE. Personal use of this material is permitted. Permission from IEEE must be obtained for all other uses, in any current or future media, including reprinting/republishing this material for advertising or promotional purposes, creating new collective works, for resale or redistribution to servers or lists, or reuse of any copyrighted component of this work in other works. 
  This work has been accepted at the 35th IEEE International Conference on Application-specific Systems, Architectures and Processors (ASAP'24).
  \end{minipage}
}

\usepackage{cite}
\usepackage{booktabs}

\ifCLASSINFOpdf
  \usepackage[pdftex]{graphicx}
\else
\fi
\usepackage{amsmath}
\usepackage{algorithmic}
\usepackage{algorithm}

\usepackage{array}

\ifCLASSOPTIONcompsoc
  \usepackage[caption=false,font=normalsize,labelfont=sf,textfont=sf]{subfig}
\else
  \usepackage[caption=false,font=footnotesize]{subfig}
\fi
\begin{document}

%
\title{Accelerating MRI Uncertainty Estimation with Mask-based Bayesian Neural Network}



%

\author{
\IEEEauthorblockN{
Zehuan Zhang\IEEEauthorrefmark{1},
Matej Genci\IEEEauthorrefmark{1},
Hongxiang Fan\IEEEauthorrefmark{1}, 
Andreas Wetscherek\IEEEauthorrefmark{2}, and
Wayne Luk\IEEEauthorrefmark{1}
}
\IEEEauthorblockA{
\IEEEauthorrefmark{1}Dept of Computing, Imperial College London, \textit{\{zehuanzhang, m.genci18, h.fan17,  w.luk\}@imperial.ac.uk}
}
\IEEEauthorblockA{
\IEEEauthorrefmark{2}Joint Dept of Physics, The Institute of Cancer Research and The Royal Marsden Hospital, London, \\
\textit{Andreas.Wetscherek@icr.ac.uk
}}
}


\maketitle

\begin{abstract}

Accurate and reliable Magnetic Resonance Imaging (MRI) analysis is particularly important for  adaptive radiotherapy, a recent medical advance capable of improving cancer diagnosis and treatment. Recent studies have shown that IVIM-NET, a deep neural network (DNN), can achieve high accuracy in MRI analysis, indicating the potential of deep learning to enhance diagnostic capabilities in healthcare. 
However, IVIM-NET does not provide calibrated uncertainty information needed for reliable and trustworthy predictions in healthcare. 
Moreover, the expensive computation and memory demands of IVIM-NET reduce hardware performance,
hindering widespread adoption in realistic scenarios.
To address these challenges, this paper proposes an algorithm–hardware co-optimization flow for high-performance and reliable MRI analysis. 
At the algorithm level, a transformation design flow is introduced to convert IVIM-NET to a mask-based Bayesian Neural Network (BayesNN), facilitating reliable and efficient uncertainty estimation. 
At the hardware level, we propose an FPGA-based accelerator with several hardware optimizations, such as mask-zero skipping and operation reordering. Experimental results demonstrate that our co-design approach can satisfy the uncertainty requirements of MRI analysis, while achieving 7.5 times and 32.5 times speedup on an Xilinx VU13P FPGA compared to GPU and CPU implementations with reduced power consumption. 
\end{abstract}


%
\IEEEpeerreviewmaketitle

\section{Introduction}\label{sec:1}

Radiotherapy has been commonly used in cancer treatment, which employs ionizing radiation to destroy cancer cells. However, the radiation also affects surrounding normal tissue, thus the precision and the dose of radiation must be carefully adjusted to reduce side effects. Traditionally, imaging and radiation treatments are conducted on separate days using different machines.
The development of MR-Linac, an advanced machine capable of imaging tumours and the related organs immediately before delivering radiation, has the potential to revolutionize cancer treatment by adaptive radiotherapy~\cite{cancers16061206} – making use of real-time imaging information of the tumour regions to improve targeted therapy while minimizing radiation-induced side eﬀects. 
Recent efforts have been made to accelerate image reconstruction for adaptive radiotherapy based on magnetic resonance-guided techniques~\cite{LECOEUR2023100484}.
However, these techniques lack support for uncertainty estimation, a critical component for clinicians in treatment planning. Accurate estimation of uncertainty helps prevent overconfident predictions, thereby improving the reliability and trustworthiness of medical decisions.

To address this issue, several probabilistic deep learning approaches~\cite{kabir2018neural} can be employed to enhance adaptive radiotherapy with uncertainty estimation.
BayesNN~\cite{jospin2022hands, wang2016towards,gal2016dropout,blundell2015weight} stands out as a highly effective approach, which has gained popularity. 
However, since BayesNN necessitates multiple forward passes to obtain results, the computational load is typically several times higher than that of DNN, posing a challenge for real-time processing which is critical for adaptive radiotherapy. 
To meet the clinical requirements and help towards wider adoption, reliable and trustworthy predictions with well-calibrated estimations and fast processing speed must be attained. Therefore, it is imperative to optimize the algorithm to fully leverage the potential of existing models for MRI analysis, and design a customized accelerator to support adaptive radiotherapy.

However, there are several challenges. First, the frequent runtime sampling essential for BayesNNs execution leads to considerable resource and latency overhead, reducing the efficiency of uncertainty estimation. Second, conventional BayesNN introduces inherent randomness into the model to compute uncertainty, and weight configurations can only be determined during runtime, which complicate the development of efficient hardware solutions. Third, BayesNN usually involves multiple sampling for each data item during inference, so weights need to be reloaded multiple times, resulting in high power consumption\cite{6757323, fan2022adaptable}.

This paper proposes a novel accelerated approach based on Bayesian neural networks, to achieve high-performance MRI analysis with uncertainty estimation. It is intended to be the first step to support uncertainty estimation in adaptive radiotherapy. An algorithm-and-hardware co-design flow is developed to endow DNN with the ability to estimate the uncertainty of predictions in real-time while adhering to low power consumption requirements. 
To eliminate the runtime sampling required by conventional BayesNNs, we adopt Masksembles~\cite{durasov2021masksembles}, an efficient mask-based BayesNN for MRI uncertainty estimation.
To facilitate the seamless adoption of Masksembles for MRI analysis, we propose a novel transformation design flow that effectively converts IVIM-NET to Masksembles-IVIM, abbreviated as uIVIM-NET.
By employing pre-defined fixed masks in Masksembles, we circumvent the inherent randomness of conventional BayesNNs to allow us to efficiently skip invalid operations for further hardware optimization.
At the hardware level, benefiting from the fact that weight configurations are determined in advance, we adopt the mask-zero skipping scheme to drop the specified weights offline. In addition, in order to avoid frequent weight loadings, we reorder the calculation and adopt the batch-level scheme, significantly reducing the power consumption. 
Our experimental results demonstrate the potential of our approach for medical applications. The proposed accelerator also achieves higher performance than existing FPGA designs, and CPU and GPU implementations.

Our contributions are summarized as follows.
\begin{itemize}
    \item A algorithm–hardware co-optimization flow that converts a DNN to a hardware-efficient mask-based BaysNN. We apply it to IVIM-NET to produce uIVIM-NET providing uncertainty information for MRI analysis.
    \item A novel customized FPGA-based accelerator for the uIVIM-NET with mask-zero skipping strategy and batch-level scheme to enhance performance and reduce power consumption.
    \item Extensive experiments are conducted to evaluate our design, which demonstrate the advantages of our approach.
\end{itemize}

\section{Background and Related Work}\label{sec:2}

\subsection{Magnetic Resonance Imaging}

MRI is a non-invasive medical technique to assess the health of patients without physical penetration into their bodies~\cite{geva2006magnetic}. MRI in medical applications works by utilizing a powerful magnetic field and radio waves to generate images of the inside of the body\cite{geva2006magnetic}. In the process of MRI, a patient is placed within strong magnetic fields. The magnetic fields cause the protons, primarily those in the abundant hydrogen atoms in the body's water and fat tissues, to respond. This process emits signals and it is possible to localize the origin of these signals and create a 3D spatial mapping of different tissues in the body. An important parameter often mentioned is a b-value\cite{le1988separation}, which specifies the strength of the diffusion sensitization. For simplification, we can think of b-value as a "scale" measurement. Larger b-values capture slow moving water molecules and smaller diffusion distances. 

The MRI-generated anatomical images undergo a comprehensive analysis to acquire a thorough understanding of the body's internal condition, which is of great importance for clinical treatment and medical research. MRI analysis is a fundamental problem in the medical field, the effectiveness of which determines the treatment outcomes of many diseases. For instance, cancer is a lethal disease with significant morbidity and mortality\cite{stern2010prevalence,rogers2015incidence}. A frequently used form of treatment is radiotherapy that utilizes ionizing radiation to eradicate malignant cells. The efficacy largely depends on delivering sufficient radiation dose to the tumour without harming vital organs. If the position of tumors can be precisely located through MRI analysis, the performance of radiotherapy would be significantly augmented. Accordingly, comprehensive MRI analysis possesses the huge potential to guarantee the accurate diagnoses and facilitate treatments\cite{bi2019artificial}.

\subsection{IVIM Model and IVIM-NET}

In the field of quantitative MRI, the intravoxel incoherent motion (IVIM) model\cite{le1988separation, le1986mr}, which is able to provide internal microscopic information, shows great potential\cite{cho2017intravoxel,zhu2017predictive,ma2018quantitative,klaassen2020pathological, le1988intravoxel, le1991measuring, wirestam1997perfusion, le2008intravoxel}. The IVIM model captures key information of internal microscopic parameters and can be used to explain signal attenuation caused by microscopic motions, which are primarily characterized as a function of the diffusion of water within tissue (diffusion) and the blood flow (perfusion).

Traditionally, to fit parameters of IVIM to observed data, the least squares method and Bayesian inference\cite{spinner2021bayesian} are used on a pixel-by-pixel basis. However, these approaches suffer from long fitting times and poor repeatability of the fitted model parameters, limiting wide clinical use. The limitations of traditional fitting methods have promoted the exploration of advanced techniques to overcome these bottlenecks.

Recently, the growing of deep learning areas has greatly contributed to the advancements in the medical field. The advent of DNN spurred the creation of the IVIM-NET, designed to estimate the parameters of the IVIM model, achieving state of the art performance both in prediction quality and speed\cite{barbieri2020deep, kaandorp2021improved}, thereby holding great promise for enhancing the clinical applicability of the IVIM model in MRI analysis.

IVIM-NET~\cite{barbieri2020deep, kaandorp2021improved} is introduced to solve the inverse IVIM problem:
\begin{equation}
\frac{S}{S_{b=0}}= fe^{-bD^*}+(1-f)e^{-bD} \label{eq1}
\end{equation}
where $S$ is the signal intensity of measurements.
A set of $S$ values are measured under different conditions determined by $b$,
where $b$ represents the b-value of measured voxels. $S_{b=0}$ is the signal intensity when b-value is 0, while $D$, $D^*$ and $f$ represent the signal attenuation caused by the Brownian motion of water molecules, the signal attenuation caused by blood flow and the fraction of incoherently flowing blood flow in the tissue, respectively.

The IVIM-NET architecture consists of 4 separate sub-networks, with each sub-network dedicated to predicting one parameter: $D$, $D^*$, $f$, and $S_{b=0}$. Each sub-network has an identical fully-connected layer architecture. Input data are normalized measurements, $S/S_{b=0}$ of voxels, and the dimension of inputs equals the number of b-values of input data.  After every hidden layer, batch normalization and a ReLU activation function are applied.

\subsection{Methods of Uncertainty Estimation}\label{sec:23}

\textbf{Bayesian Neural Network (BayesNN).}
For BayesNNs, weights are treated as probability distributions instead of point values to predict the posterior of outputs. This allows estimating uncertainty by quantifying the distribution of possible outputs for a given input, rather than just a single point estimate. The commonly employed approaches are Markov Chain Monte Carlo method\cite{hastings1970monte} and Variational Inference \cite{blei2017variational}. 
The Markov Chain Monte Carlo method can be considered the best available solution to sample from exact posterior distributions, but the substantial amount of operations is prohibitively expensive for most deep learning models\cite{jospin2022hands}, rendering it unpopular. 
Variational Inference demonstrates superior scalability. The core idea is the use of another distribution to approximate the posterior. Gaussian distributions are commonly utilized as proxy distributions to finish the inference\cite{hernandez2015probabilistic,blundell2015weight}. An alternative to directly estimating model parameters is to
approximate inference from multiple predictions of the
model, which saves computational overhead. In this approach, the Monte Carlo Dropout (MC-Dropout) method~\cite{gal2016dropout} which utilizes Bernoulli distributions is popular since it does not require large modifications to existing network architectures. But it often estimates uncertainty with lower quality~\cite{gustafsson2020evaluating}. Most methods require the utilization of specific distributions to finish samplings, thereby introducing inherent randomness, which can make hardware design difficult.

\textbf{Ensemble method.}
The ensemble method for uncertainty estimation in deep learning involves using multiple models to make predictions, and then aggregating the predictions to estimate uncertainty with the variance of the individual predictions serving as a measure of uncertainty. This can be done by combining the outputs of multiple models, or by training an ensemble of models to make predictions. The Deep Ensembles method\cite{lakshminarayanan2017simple} is a kind of this methods, and is able to achieve well-calibrated uncertainty estimation. But ensemble methods typically require heavy operational costs due to implementing a large set of networks.

\textbf{Masksembles.} MC-Dropout and Deep Ensembles are popular approaches for uncertainty estimation, while there is a trade-off between the algorithm performance and computing costs between these methods~\cite{durasov2021masksembles}. Masksembles~\cite{durasov2021masksembles} is proposed to combine the advantages, and can be utilized to generate models capable of estimating uncertainty, which are defined as mask-based BayesNN in this paper. It generates a set of less correlated masks in advance, which keep fixed during training and inference as well. The masks are followed after the fully-connected (FC) layer or feature maps to keep or drop the corresponding neurons or channels. During inference, the masks are also applied. For each input, a set of sampling results are obtained to calculate predictions and uncertainty. Since the masks generated are less correlated, the quality of estimated uncertainty is comparable to Deep Ensembles. As a result, it improves performance while maintaining low computing costs. For more details about the ways to generate the masks, please refer to the work\cite{durasov2021masksembles}.

\subsection{BayesNN Accelerators}

The development of customized BayesNN accelerators has attracted much attention\cite{cai2018vibnn,awano2020bynqnet,fan2021high,fan2022accelerating,fan2022enabling, ferianc2021optimizing,fan2022fpga}. The work\cite{cai2018vibnn} is the first to accelerate BayesNNs based on Variational Inference, which elaborated on ways to generate random numbers in detail. BYNQNet\cite{awano2020bynqnet} exploits the sampling-free method and implements the model on the PYNQ-Z1 board. The approach adopts moment propagation for inference at a low hardware cost. ASBNN\cite{fujiwara2021asbnn} explores the relationship among multiple forward passes to achieve approximate calculations. The methods in \cite{fan2021high}\cite{fan2022accelerating} involve thorough explorations on the structured sparsity of Monte Carlo Dropout-based convolutional BayesNNs, and designed FPGA-based hardware accelerators.
\cite{fan2023monte} combined Monte Carlo Dropout and Multi-Exit methods and designed accelerators with spatial-temporal mapping strategies.
Nonetheless, previous designs have not been applied to support real-time MRI analysis.

\section{Algorithm–hardware co-optimization flow}

\begin{figure*}[htbp]
\centerline{\includegraphics[width=7in]{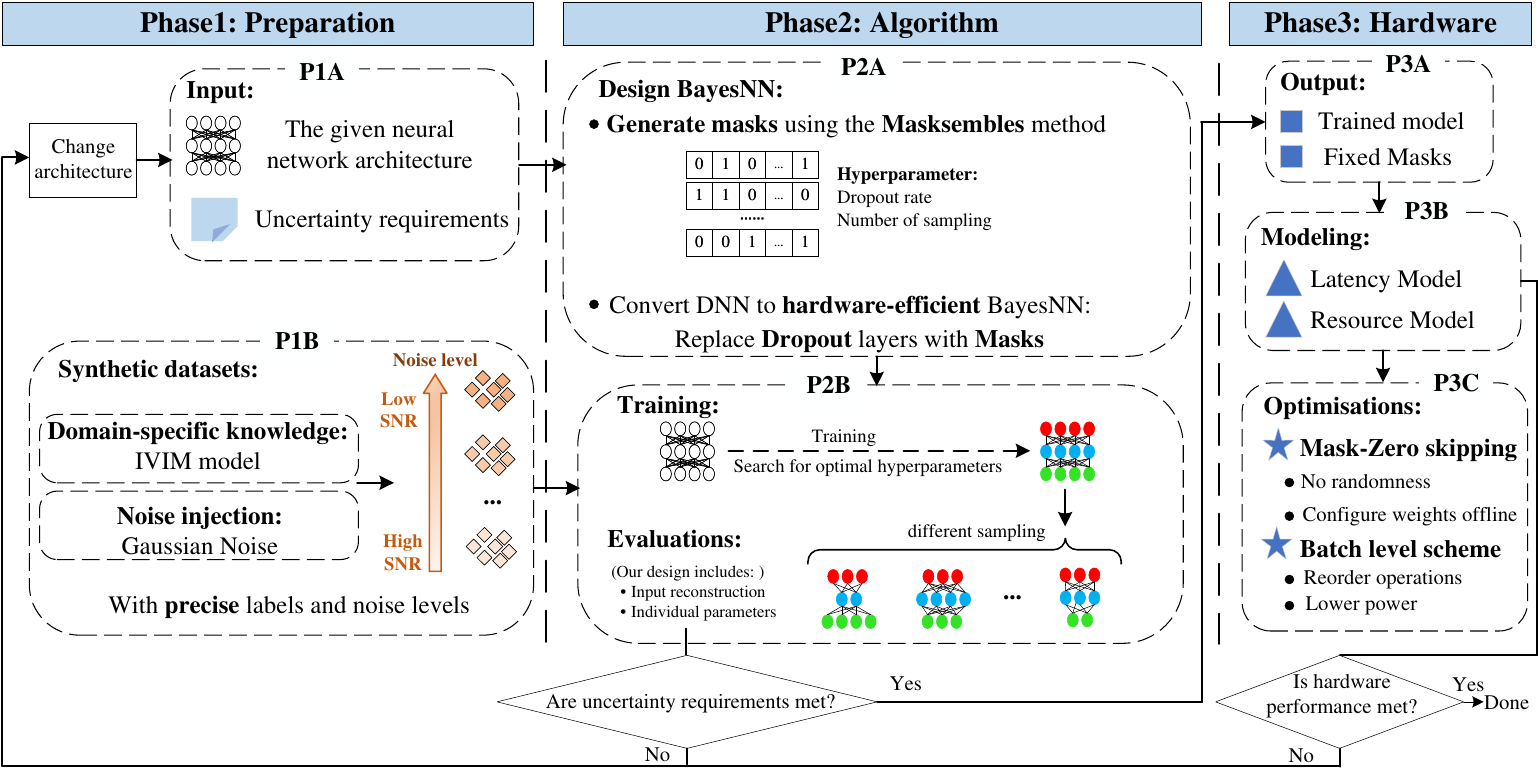}}
\caption{The algorithm–hardware co-optimization flow from DNN to mask-based BayesNN with hardware optimization}
\label{fig:1}
\end{figure*}

An overview of the algorithm–hardware co-optimization flow is shown in Fig.~\ref{fig:1}. It is proposed to convert DNN to mask-based BayesNN characterized by hardware-efficient architectures, and to provide hardware optimization strategies to realize efficient model deployment on FPGA.

\textbf{Phase 1: Preparation.}
In the first phase, a neural network architecture should be given. Theoretically, most main-stream networks equipped with dropout\cite{srivastava2014dropout} layers, which are the popular methods for regularization, are all compatible. Also, uncertainty requirements tailored to the situation and particular constraints are formulated. The uncertainty requirements serve as a basis to assess the uncertainty quality as well.

In addition, synthetic datasets are required in the workflow. Typically, models are trained on collected real datasets that have been manually labeled. While this approach is effective for evaluating accuracy, it presents difficulties in assessing uncertainty estimates due to the absence of ground truths of uncertainty for the collected data. To resolve this issue, the utilization of synthetic data is a must. A multitude of datasets are simulated based on domain-specific knowledge. Furthermore, different levels of noise are also injected into simulated data in accordance with predefined uncertainty requirements to generate distinct synthetic data, each representing a scenario. More simulated situations enable a more comprehensive evaluation of the network's performance across diverse scenarios. In this way, precise labels and noise levels can be obtained easily for synthetic data, which serves as a viable solution to the issue of collecting measured data and ground truths.

\textbf{Phase 2: Algorithm.}
The second phase processes model design, training and evaluations. To convert the given architecture to a BayesNN, the Masksembles approach is selected for this purpose, since it covers a range of ensemble-like models of which Monte Carlo Dropout\cite{gal2016dropout} and Deep Ensembles\cite{lakshminarayanan2017simple} are extreme examples\cite{durasov2021masksembles} as stated in \ref{sec:23}. The hyperparameter settings of the Masksembles can also be regulated to ensure high-quality uncertainty estimates. Moreover, it is a plug-in module that can be directly inserted into an existing neural network, requiring only minor modifications to the network. Hence, it is characterized by its general applicability. In addition, since the masks are fixed, the position of neurons to be retained or dropped can be determined explicitly, thereby eliminating the randomness during inference, making it more efficient for hardware design.

Then, the model is trained on synthetic datasets. Given that the dropped positions are predetermined, it works like an enhanced version of conventional dropout techniques. A grid search is conducted for the dropout rate ranging from 0.1 to 0.9 with a step size of 0.1, and the sampling number is varied among 4, 8, 16, 32, 64 to find the optimal hyperparameters.

After training, synthetic data with ground truths are used for evaluations. Evaluation results show whether the network satisfies uncertainty requirements across all situations, or whether it exhibits subpar performance in certain situations. These results could provide valuable insights for practical applications.

If all uncertainty requirements are satisfied, the design flow continues to Phase 3 for hardware design; otherwise, it implements iterative improvements, such as changing the given model architecture, and then restarting the flow.

\textbf{Phase 3: Hardware.}
In the third phase, the trained model and the fixed mask generated in Phase 2 are obtained. We model latency and resource consumption, and concentrate on efficient hardware architecture design.

For latency and resource models, they mainly depend on the size of the network, the pipeline design and the DSP resource consumed. We propose simple analytic formulations for their estimation. If high performance is desirable, we can map the network onto an FPGA with efficient hardware optimizations.

The model obtained contains hardware-efficient characteristics, which are exploited to address pain points of BayesNN in hardware design. We propose mask-zero skipping strategies to configure weights offline, eliminating randomness completely. We also reorder operations to avoid frequent weight loading, significantly reducing power consumption. Further details are elaborated in Section~\ref{sec:5}.

\section{Algorithm Design of uIVIM-NET}

\begin{figure}[ht]
\centerline{\includegraphics[width=3.4in]{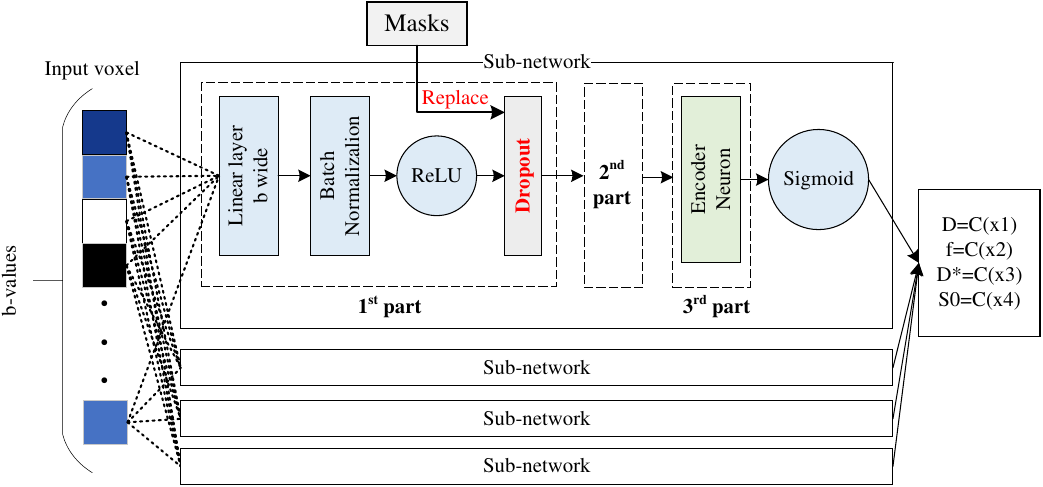}}
\caption{Conversion from IVIM-NET to uIVIM-NET }
\label{fig:2}
\end{figure}

Fig.~\ref{fig:2} illustrates the architecture of IVIM-NET, consisting of 4 identical separate sub-networks, and each sub-network mainly consists of 3 parts. In the first part, one linear layer is followed by batch normalization, the ReLU function, and the dropout layer. The second part is the same as the first part. The third part is only one linear layer, called the encoder. The outputs of the encoder are fed to the Sigmoid function to be the outputs of this sub-network. Finally, the conversion function, denoted as $\mathrm{C}(\cdot)$, transforms the sub-network outputs to the corresponding parameters as results. The width of linear layers equals to the number of b-values of input voxels. Regarding uncertainty requirements, it is expected that output uncertainty shrinks with less noise.

In order to produce synthetic data, the physical equation (\ref{eq1}) is exploited to simulate data. The ranges of $S_0$, $D^*$, $D$, and $f$ are set according to real scenarios, then these parameters are randomly specified to calculate signal density $S$ for each data point. Afterwards, noise is injected to corrupt clean data, which frequently occurs in real collected data. Gaussian noise with mean 0 and standard deviation $S_0/SNR$ is added with different signal-to-noise ratios (SNR) which determine the noise levels simulating different scenarios. As a result, synthetic data with different noise levels are produced.

To create uIVIM-NET, the dropout blocks are replaced by masks given by the Masksembles, and these are enabled during training and inference. For training, the loss function is in line with that of IVIM-NET. Specifically, each network is responsible for estimating a specific parameter that can be utilized to reconstruct inputs. The loss is calculated as the mean-square error (MSE) between the input and the reconstructed input derived using equation (\ref{eq1}).

In the evaluation stage, for every input, the network is evaluated multiple times to obtain a set of sampling of predictions of individual parameters and reconstruction. The mean is the final prediction value, and the standard deviation provides a measure of the associated uncertainty.

\section{Hardware Design}\label{sec:5}

\begin{figure}[h]
\centerline{\includegraphics[width=0.42\textwidth]{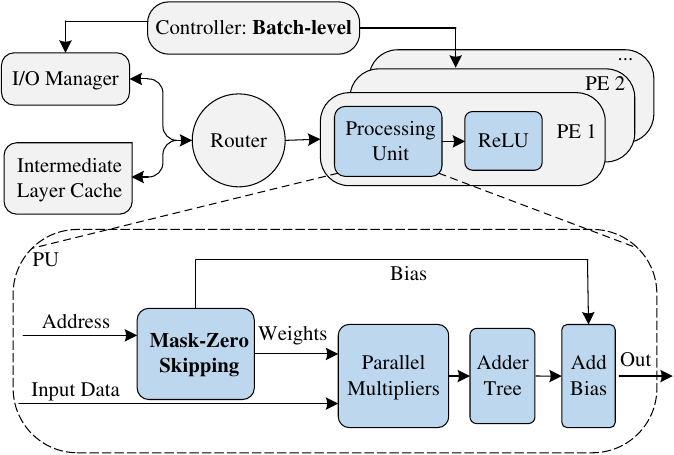}}
\caption{Hardware architecture}
\label{fig:4}
\end{figure}

In Phase 3, we develop a parallel and pipelined uIVIM-NET accelerator by tailoring the accelerator architecture specifically to the model. Fig.~\ref{fig:4} gives an overview architecture consisting of an I/O manager, an intermediate layer cache, multiple identical processing elements (PEs), and a controller. The I/O manager serves as a repository for input data and outputs from PEs, while the intermediate layer cache is utilized to temporarily store intermediate results. The controller manages the internal state of the accelerator, overseeing the progress of the computation, coordinating the sourcing and storage of data, and determining when to retrieve a sample.

\subsection{Analysis of Parallelism}
The uIVIM-NET features several parallelism opportunities. First, input parallelism can be achieved by processing multiple elements in the input data concurrently, especially when processing all elements simultaneously in one voxel. Second, output parallelism is attainable by designing different PEs to handle independent output neurons in FC layers. Third, sub-network parallelism can be exploited by creating parallel processing blocks for each of 4 independent separate sub-networks, requiring more DSP resources. Fourth, sample parallelism can be achieved by processing all sampling obtained from evaluating each sub-network multiple times simultaneously. In summary, our design prioritizes input and output parallelism due to resource constraints, while sub-network parallelism and sample parallelism are not picked.  It is possible to process the latter serially and we adopt optimizations based on the batch-level scheme.

\subsection{Memory}

The I/O manager stores input data and outputs from PEs. It is implemented using BRAM blocks, so the entirety of the inputs should be held in on-chip memory. If the amount of voxels exceeds the limit, it is possible to store input data in batches since all the voxels are independent.

The intermediate layer cache is used to store temporary results. Since the model is computed layer by layer, the results of early layers are stored in the cache. Moreover, if the size of a linear layer is larger than the number of processing elements, the intermediate layer cache stores partial results of computing linear layers serially as well.

\subsection{Processing Module}

Processing modules are responsible for calculating the outputs of layers in the model. As shown in Fig.~\ref{fig:4}, input data are passed into processing units (PU) first, which includes a block of parallel multipliers followed by a tree of adders to finish dot product calculations, and then bias is added. The activation function performs ReLU. The processing modules are logically organized into identical PEs, with each PE dedicated to computing for a single output neuron, all operating in parallel. Each forward pass is computed layer by layer. Different layers in multiple forward passes share PEs. 

\begin{figure}[htbp]
\centerline{\includegraphics[width=3.4in]{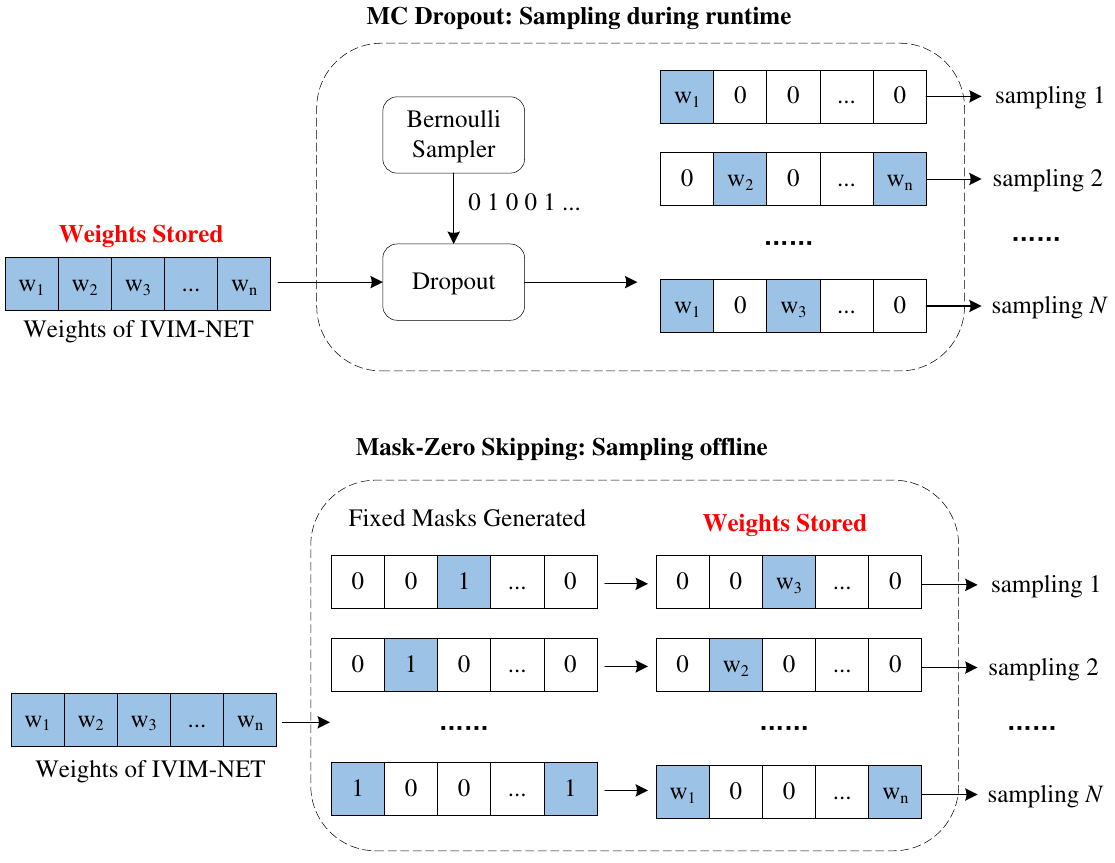}}
\caption{MC Dropout scheme and mask-zero skipping scheme. MC Dropout scheme dynamically drops weights during runtime. Mask-zero skipping scheme stores dropped weights offline.}
\label{fig:5}
\end{figure}

For each PU in PEs, there is a dedicated memory block to store weights and biases of the uIVIM-NET. Fig.~\ref{fig:5} shows the comparison of the previous common scheme and our scheme. In previous accelerators\cite{fan2021high,fan2022accelerating} that target for BayesNN based on MC-Dropout, it requires sampling weights following the Bernoulli distribution. In order to achieve sampling randomly and drop corresponding weights, the weights have to be determined during runtime by the Bernoulli Sampler module, and the Dropout module drops corresponding neurons. In this way, it increases the consumed resources, and also increases the latency and power due to more incurred operations. In order to overcome this bottleneck, we propose a mask-zero skipping storage strategy. For the uIVIM-NET, since the dropped positions of weights have been predetermined and are fixed, it is allowed to only store weights which are not dropped, avoiding the need for other modules. Moreover, it is a must to keep some copies, the number of which equals the number of sampling.

To achieve maximum performance, we adopt fine-grained pipelining within PUs to maximize throughput. We insert $R_A$ and $R_M$ internal pipeline registers to every adder and every multiplier. Implementing internal pipeline registers minimizes the path between registers, thus allowing for an increase in clock speeds, which in turn speeds up the computation. Pipeline stages are independent; therefore, every adder and multiplier can start processing a new value every clock cycle, despite the latency of every computation being \textgreater 1. Therefore, if the number of PEs is $N_{PE}$, the number of b-values is $N_b$, which is also the dimension of inputs, and the adder tree is $L$ levels deep. The total latency of a PU is the time it takes to perform multiplication, evaluate the adder tree, accumulate the result of $\lceil N_b/N_{PE}\rceil$ parts over time, and add bias:
\begin{equation}
\begin{aligned}
Latency\ of\ PU &= R_M + L \times R_A+(\lceil N_b/N_{PE}\rceil-1)+R_A \\
&=R_M + R_A (L+1) + \lceil N_b/N_{PE}\rceil - 1
\end{aligned}
\label{eq2}
\end{equation}

\subsection{Controller}
To coordinate all modules within the accelerator, the controller uses an internal state machine to dispatch control signals to evaluate each sub-network. 

\begin{figure}[htbp]
\centerline{\includegraphics[width=3.4in]{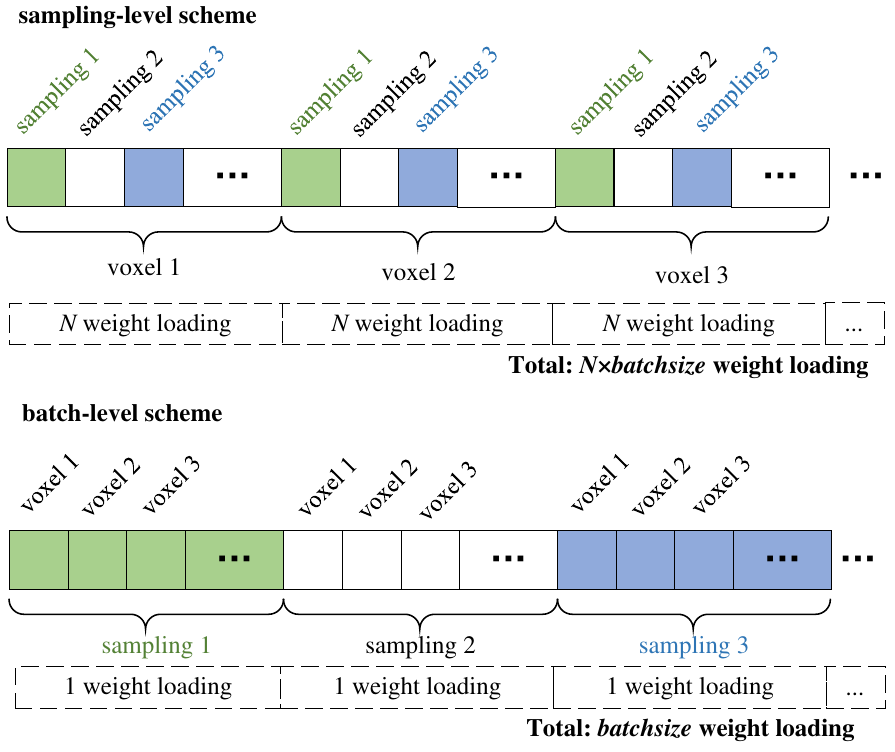}}
\caption{Sampling-level scheme and batch-level scheme}
\label{fig:6}
\end{figure}

During the inference stage, each sub-network requires multiple forward passes to obtain different sampling. Typically, processing sampling serially is adopted as the operation order, denoted as the sampling-level scheme, as illustrated in Fig.~\ref{fig:6}. In this order, for each voxel, weights of each sampling must be loaded multiple times to complete sampling, which unavoidably incurs frequent loading operations, increasing the power significantly\cite{6757323}. To address this issue, we introduce a new batch-level operational order. As described previously, the generated masks are fixed, which means multiple weight sampling should not change for all the voxels. Hence, there is no need to reload each sampling weight repeatedly for each voxel since the same weight configurations can be implemented many times. 

Therefore, for the batch-level scheme, only the weights of one sampling are loaded for evaluations of the whole batch of voxels. Then, once this batch of evaluations is complete, the weights of the next sampling are loaded for evaluations, continuing until all samplings are processed for all voxels in this batch. Afterwards, a new cycle of evaluation begins with the next batch of voxels. Using this scheme, weights of each sampling are only loaded once per batch of voxels. If the number of sampling is $N$, for each batch, the sampling-level scheme requires $N \times batchsize$ weight loadings while the batch-level scheme requires $N$ weight loadings. Our scheme reduces the weight loading operations by $batchsize$ times, thereby decreasing power consumption.

\section{Evaluation}\label{Cha:5}

\subsection{Experimental Setup}

We use an Intel Xeon Silver 4110 as our CPU platform and an Nvidia GeForce GTX 1080 Ti as our GPU platform to run the uIVIM-NET using Pytorch (1.10.0) framework for the software baseline. Signals are generated using the equation (\ref{eq1}) by drawing $S_0$, $D^*$, $D$, and $f$ randomly from reasonable ranges in real cases according to domain experts, with added Gaussian noise. Synthetic datasets with 5 different levels of noise, corresponding to SNR values of 5, 15, 20, 30, and 50, were generated, with each dataset containing 10,000 synthetic data. For each data, $S/S_0$ is calculated as inputs of the model.

Our accelerator is designed using Xilinx Vivado 2021.2 written in Verilog, targeting the Xilinx VU13P device, running at 250MHz. The quantization scheme is 16-bit fixed-point representation with 4 integer bits. The simulation results, resource utilization and power consumption after implementation on the Vivado tool are reported. The accelerator features 32 PEs, with each PE capable of processing voxels up to 128 elements, which is enough to support real scenarios, a published IVIM dataset\cite{gurney2018comparison,klaassen2018evaluation,klaassen2018repeatability} with 104 b-values. On chip, 20k voxels are stored with a batch size of 64 and a sampling number of 4.

\subsection{Algorithm Performance}

\begin{figure}[htbp]
\centerline{\includegraphics[width=3.4in]{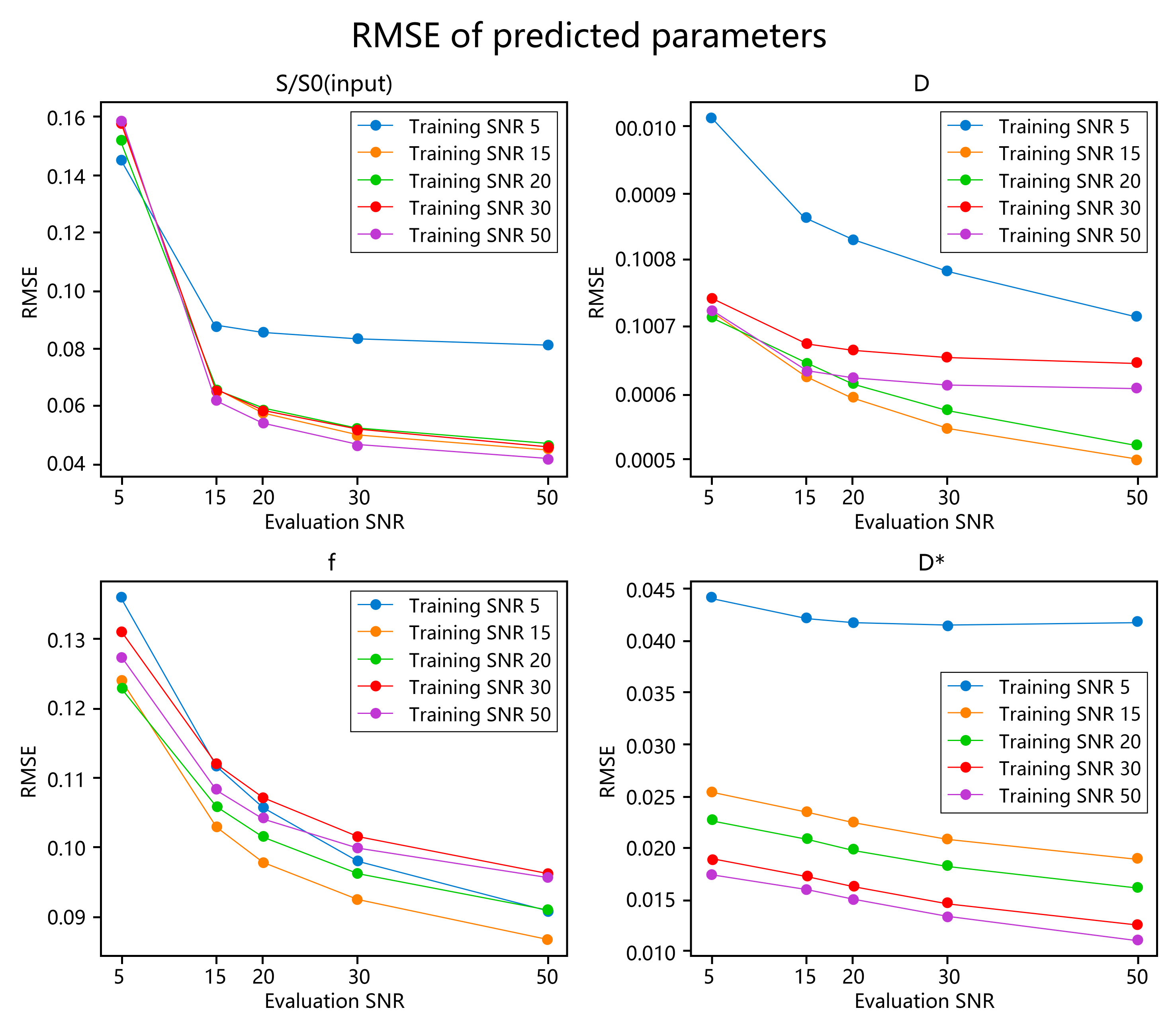}}
\caption{RMSE of predicted parameters}
\label{fig:9}
\end{figure}

\begin{figure}[htbp]
\centerline{\includegraphics[width=3.4in]{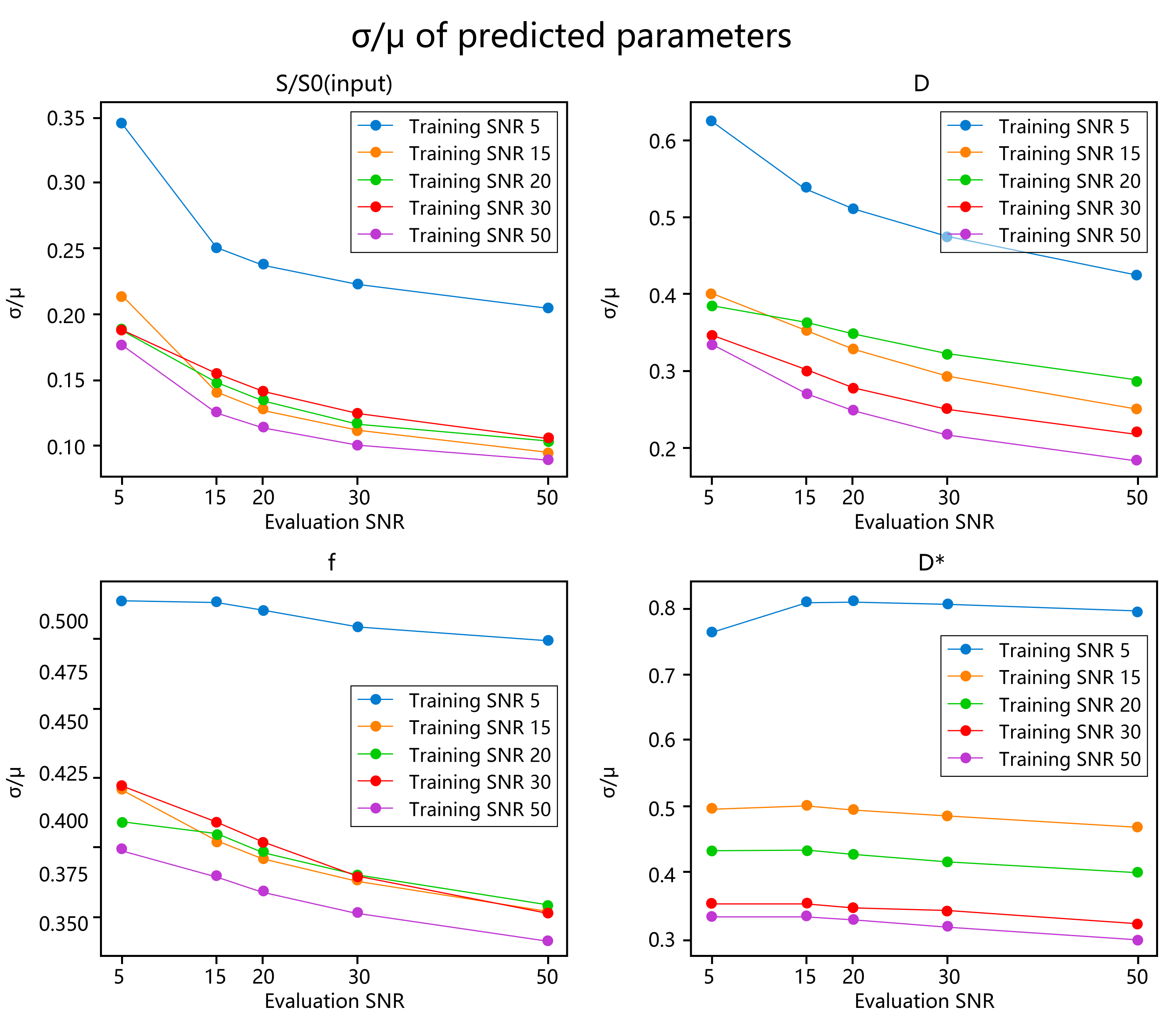}}
\caption{Uncertainty of predicted parameters}
\label{fig:10}
\end{figure}

Root means squared error (RMSE) between, the reconstruction and predicted individual IVIM parameters, and their respective ground truth values is calculated to assess the overall accuracy. The evaluation results of different SNR values are plotted in Fig.~\ref{fig:9}. Furthermore, the standard deviation (std) divided by the mean of samples is used to assess the uncertainty for each data. This metric measures the relative variance. The evaluation results of different SNR values are also plotted in Fig.~\ref{fig:10}. 

The figures show that less noise in input voxels (evaluation SNR is higher) leads to smaller RMSE (higher accuracy) and low uncertainty (more confident), and typically, networks trained with less noisy data tend to exhibit higher levels of accuracy and confidence, which is in line with expectations.

The results demonstrate the effectiveness of the framework in successfully converting the existing IVIM-NET to uIVIM-NET empowered with the ability to estimate the uncertainty of predictions. The findings on synthetic data can be a valuable reference to provide guidance in real medical scenarios, thus exhibiting significant potential to enhance MRI analysis. For instance, clinicians are able to set numerical thresholds to determine diagnosis with high uncertainty based on the experimental results on synthetic data, and further adopt more comprehensive medical examinations to treat patients.

\subsection{Hardware Performance}

\begin{table*}[h]
\begin{center}  
\caption{Comparison with state-of-the-art designs\\(The speed metric is not included as the accelerated neural network models are different)}  
\label{table:2} 
\begin{tabular}{l|l|l|l|l|l}
\toprule
 & \textbf{ASPLOS'18\cite{cai2018vibnn}}     & \textbf{DATE'20\cite{awano2020bynqnet}}    & \textbf{DAC'21\cite{fan2021high}}    & \textbf{TPDS'22\cite{fan2022accelerating}}    & \textbf{Ours}    \\ \hline
\textbf{Platform}     & Altera Cyclone V & Xilinx Zynq XC7Z020 & Arria 10 GX1150 & Arria 10 GX1150 & Xilinx VU13P \\ \hline
\textbf{Frequency}    & 213MHz    & 200MHz    & 225MHz      & 220MHz    & 250MHz       \\ \hline
\textbf{Power}(W)           & 6.11            & 2.76       & 45.00    & 43.60  &11.78            \\ \hline
\textbf{Neural Network Model}        & Bayes-FC          & Bayes-FC          & Bayes-VGG11    & Bayes-VGG11        & Mask-based Bayes-FC               \\ \hline
\textbf{Technology}      & 28nm       & 28nm     & 20nm     & 20nm      & 16nm              \\ \hline
\textbf{Energy Efficiency}(GOP/s/W) & 9.75        & 8.77         & 11.9       & 19.6   & 20.31       \\ \bottomrule
\end{tabular}
\end{center}  
\vspace{-3.0mm}
\end{table*}

\paragraph{Comparison with existing designs}

To demonstrate the benefits of our converted algorithm and customized hardware architecture and optimizations as a whole, we make comparisons against the existing designs in Table \ref{table:2}. Previous work~\cite{cai2018vibnn,awano2020bynqnet,fan2021high,fan2022accelerating} accelerated BayesNNs, but they were not evaluated for medical analysis. As these designs were evaluated on different BayesNNs, it is unfair to compare them in terms of speed unless the same network is executed. 
Therefore, we choose the energy efficiency, throughput per watt consumed, as metrics. We quote the hardware performance from the original papers. Table \ref{table:2} shows that our design achieves more than $2 \times$ energy efficiency compared with the work \cite{cai2018vibnn} and \cite{awano2020bynqnet} which accelerate BayesNN only consisted of FC layers similar to ours, indicating significant improvements. The accelerators\cite{fan2021high,fan2022accelerating} are optimized for convolutional BayesNN requiring denser operations, resulting in higher power consumption. Compared with those, our design also exhibits higher energy efficiency, demonstrating the effectiveness of the design flow.

The advantages of low power and high energy efficiency in our design could be attributed to the hardware optimizations. Firstly in previous work~\cite{cai2018vibnn,fan2021high,fan2022accelerating}, random number generators are designed to determine samplings and dropout operations are implemented during runtime, while the omission of the Sampler and Dropout modules in our architecture results in a significant reduction in power. Secondly, by utilizing a batch-level scheme, the frequent loading of weights is avoided, thus leading to low power consumption.

To further showcase the benefits in comparison to existing work, an estimated comparison is presented. Firstly, our approach is a co-design framework. The adoption of Masksembles enhances the software performance and yields advantages in hardware design as well. It is worth noting that previous work~\cite{cai2018vibnn,fan2021high,fan2022accelerating} only focused on accelerating MC-Dropout BayesNN, so the potential from algorithmic aspect is unexplored. Secondly, hardware optimizations applied to algorithmic architecture generated from the conversion flow reduce power consumption as shown and discussed before. Thirdly, the whole flow of our methodology is more general and can be extended to other mainstream neural networks, thus showcasing significant potential for widespread applications.

\paragraph{Comparison with CPU and GPU}
A hardware performance comparison of our FPGA design with both CPU and GPU implementations is shown in Table \ref{table:1}.

\begin{table}[htb]   
\begin{center}   
\caption{Comparison against CPU and GPU implementations}  
\label{table:1} 
\begin{tabular}{l|ll|l}
\toprule
          & \multicolumn{1}{l|}{\textbf{CPU}}       & \textbf{GPU}       & \textbf{Ours}    
\\ \hline

\textbf{Platform} & \multicolumn{1}{l|}{\begin{tabular}[c]{@{}l@{}}Intel Xeon 
           \\ Silver 4110\end{tabular}} 
           & \begin{tabular}[c]{@{}l@{}}GeForce GTX \\ 
              1080 Ti\end{tabular} 
         & Xilinx VU13P  \\ 
\hline     
\textbf{Compiler}     & \multicolumn{2}{l|}{     Pytorch 1.10.0}     & Vivado 2021.2 \\ 
\hline
\textbf{Frequency}    & \multicolumn{1}{l|}{2.10GHz}   & 1.48GHz     & 250MHz \\ 
\hline
\textbf{Technology}   & \multicolumn{1}{l|}{14nm}       & 16nm       & 16nm              
\\ \hline
\begin{tabular}[c]{@{}l@{}}\textbf{Latency} 
\\ {(}ms/Batch{)}\end{tabular} 
& \multicolumn{1}{l|}{9.1}      
& 2.1       
& 0.28  
\\ \hline
\textbf{Power}(W)      
& \multicolumn{1}{l|}{30}       
& 54             
& 11.78      
\\ \hline
\begin{tabular}[c]{@{}l@{}}\textbf{Energy Cost} 
\\ {(}mJ/Batch{)}\end{tabular} 
& \multicolumn{1}{l|}{273}      
& 113.4           
& 3.3 
\\ \bottomrule
\end{tabular}   
\end{center}  
\end{table}

Each batch takes our FPGA accelerator 0.28ms on average. The acceleration performance outperforms that of the GPU and CPU by a factor of 7.5 and 32.5, respectively, while using only a fraction of the power. To further illustrate this advantage, we calculated the energy cost of each batch of voxels across different platforms. As indicated in Table \ref{table:1}, the proposed design demonstrates substantial energy cost savings per batch, with reductions of approximately $34.4 \times$ and $82.8 \times$ when compared to the GPU and CPU versions, respectively. To support uncertainty estimation in adaptive radiotherapy, the processing speed should achieve $0.8ms/batch$. The fast speed of our accelerator meets the real-time requirement, indicating considerable promise for practical applications.

\paragraph{Resource Utilization}

\begin{figure}[htbp]
\centerline{\includegraphics[width=0.44\textwidth]{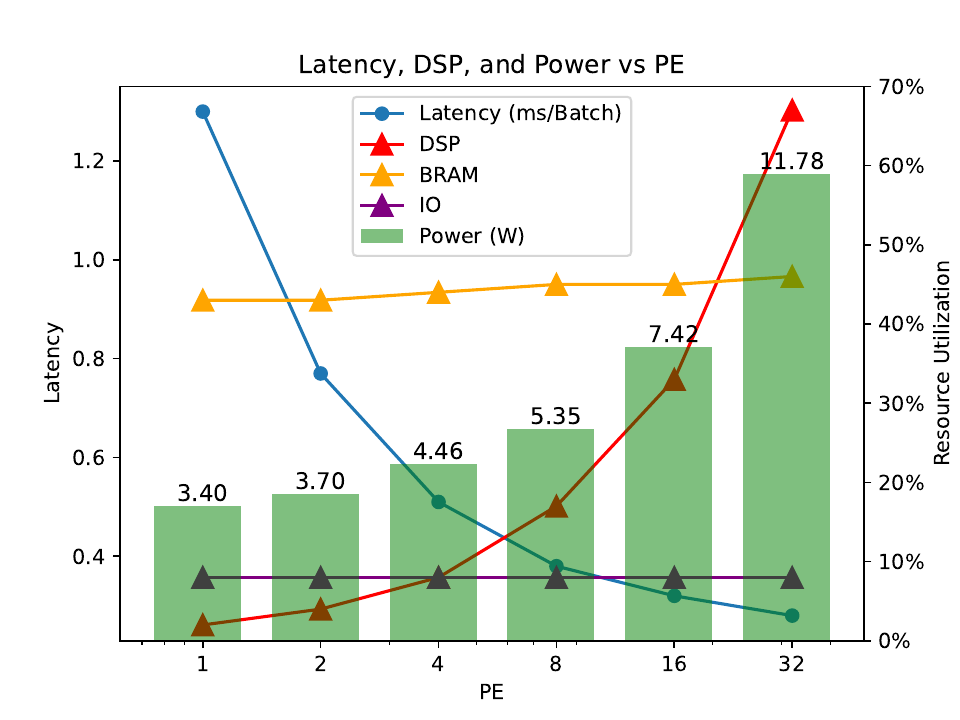}}
\caption{Relationship between resource utilization
and performance}
\label{fig:12}
\end{figure}

More experiments are conducted to evaluate resource utilization and performance of the accelerator with varying numbers of PEs. The results, shown in Fig.~\ref{fig:12}, confirm the main limiting factor of the speed of our design is the amount of available DSP resources. Specifically, when 32 PEs are deployed, the accelerator consumes 67\% of all available DSPs with 11.78W power.

The relationship between consumed resources and processing speed can also be observed. The utilization of BRAM and IO resources remains relatively constant, as the consumption of BRAM resources primarily depends on the storage of voxels and model weights, while the consumption of IO resources is primarily influenced by the ways to access the outcomes. The number of DSPs for each PE is constant, so DSP resources consumed are proportional to the number of PEs. The processing speed can be estimated based on equation (\ref{eq2}), which matches the practical results. With higher parallelism, the accelerator achieves faster speed but also increases power consumption and resource utilization. The relationship presents a trade-off. The parallelism can be determined according to resources available on chip and performance requirements.

\section{Conclusion}
This paper presents an algorithm and hardware co-design approach applicable to existing DNNs. This methodology is adopted to optimize medical imaging models to estimate uncertainty and to customize an accelerator for fast MRI analysis. Experiments demonstrate the effectiveness in various scenarios, which provide useful insights for applications. Our customized accelerator also exhibits remarkable computational speed and energy efficiency, outperforming both CPUs and GPUs as well as existing FPGA designs. 

On the basis of the current research, further exploration holds significant potential as well.
It is promising to adopt the proposed accelerator to support image-guided treatment, which would involve integrating with other functions such as dose calculation to provide a real-time system for adaptive radiotherapy\cite{voss2020towards}.

Furthermore, apart from MRI analysis, uncertainty information is also helpful for other applications such as intelligent robots and autonomous driving. 
The active agents and controller need to make decisions based on incomplete knowledge, and the assumption that the inference situation has the same distribution as training is often invalid in many real scenarios. 
It is believed that with uncertainty information, more robust performance could be attained. 
Our proposed framework can be extended to cover these applications.

\section*{Acknowledgement}
The support of UK EPSRC (grant number EP/X036006/1, EP/P010040/1, EP/V028251/1 and EP/S030069/1), AMD and Intel is gratefully acknowledged.

\clearpage
\bibliographystyle{IEEEtran}
\bibliography{bibliography}

\end{document}